\documentstyle[11pt,epsbox]{article}
\begin{document}
\pagestyle{empty}
\begin{center}
\begin{Large}
{\bf Spin-Wave Excitations of
Half-Filled Kondo Lattice Model} \\
\end{Large}
\hspace*{1.5mm} \\
{\bf Yasuhiro Matsushita}\footnote{
e-mail: matusita@grad.ap.kagu.sut.ac.jp}, {\bf Ryousuke Shiina}
 {\bf and Chikara Ishii} \\ 
{\it Department of Physics, Science University of Tokyo, \\
1-3 Kagurazaka, Shinjuku-ku, Tokyo 162}
\end{center}

\vspace{20pt}
\begin{quote}
\begin{footnotesize}
 The spin excitations in the antiferromagnetic phase of 
half-filled Kondo lattice model 
are studied by means of the decoupling approximation for spin 
Green's function. 
The spin-wave spectrum is calculated as a function of Kondo 
coupling, and this is used to calculate the thermodynamic quantities 
at low temperatures. 
The N\'{e}el temperature of the form 
  $T_{\rm N}=0.087{J}^2\ln{(1/J)}$ 
is obtained for the 3-dimensional case 
in the weak-coupling limit. 
It is pointed out that the ratio of the spin-wave velocity 
to the N\'{e}el temperature $v_s/T_{\rm N}$ is enhanced as the 
Kondo coupling becomes small, reflecting the long-range nature 
of effective interactions between localized spins. 
\end{footnotesize}
\end{quote}

\noindent
\setcounter{page}{1}
\pagestyle{plain}

The electronic states of heavy-fermion compounds 
have been extensively studied to understand 
a number of their anomalous properties due to strong correlations. 
The Kondo lattice model (KLM) and 
the periodic Anderson model (PAM) have been considered to give a 
microscopic basis to describe 
the characteristic features of these systems. 
It has been expected that KLM and PAM exhibit 
various types of magnetic long-range order 
due to effective RKKY interaction between localized spins. 
In particular, the properties of 
antiferromagnetically ordered (AFO) phase realized 
in the half-filled system have been studied 
by the Gutzwiller variational method \cite{faz}\cite{Zs} 
and the slave-boson mean-field approximation \cite{dor}\cite{sun} 
in connection with the magnetic instability of 
heavy-fermion semiconductors \cite{sn}. 
However, the spin excitations and the resulting thermodynamic properties 
in AFO phase have not been fully understood, 
because the low-lying excitations are closely related to 
the non-local fluctuations neglected in these approximations. 
In the present paper, we are going to investigate these spin excitations 
in the AFO phase of KLM, making use of  the spin Green's functions 
in the decoupling approximation similar to 
the Tyabrikov method for the Heisenberg model \cite{bog}. 
The spin-wave excitation spectrum and the N\'{e}el temperature 
are determined from these Green's functions. 

The KLM with the nearest-neighbor hopping is described 
by the Hamiltonian
$$
 H=-t\sum_{<ij>\sigma} c_{i\sigma}^\dagger c_{j\sigma}
+J \sum_i \vec{\sigma}_i \cdot \vec{S}_i, 
\eqno{(1)}
$$
where $c_{i\sigma}^\dagger$ and $c_{i\sigma}$ represent 
the creation and annihilation operators of a conduction electron 
in Wannier state at the i-th site with spin ${\sigma}$, respectively. 
The spin operator of conduction electron $ \vec{\sigma}_i$ 
is expressed in the form
$  \vec{\sigma}_i=1/2 \sum_{\sigma\sigma'} c_{i\sigma}^\dagger
     \vec{\tau}_{\sigma\sigma'} c_{i\sigma'}, $
where $ \vec{\tau} $ is the Pauli matrix. 
In the following, we set $t=1$, measuring all energies 
in the unit of $t$. 

In order to describe AFO phase in this model, 
we introduce the staggered magnetizations 
of localized spin and conduction electron, respectively : 
$$\langle S_i^z \rangle=\left\{
\begin{array}{rl}
m_s&\quad\mbox{for A sublattice}\\
-m_s&\quad\mbox{for B sublattice,} 
\end{array} \right.
$$ 
$$\langle \sigma_i^z \rangle=\left\{
\begin{array}{rl}
-m_c&\quad\mbox{for A sublattice}\\
m_c&\quad\mbox{for B sublattice.} 
\end{array} \right.
$$ 
Then the Hamiltonian (1) is devived by 
the mean-field part $H_{0}$ and the residual-interaction part 
$H_{1}=H-H_{0}$. 
The $H_{0}$ describes 
the conduction band in the staggered field, 
and can be regarded as the quasi-particle bands 
with spin-density-wave (SDW) modulation
by means of the canonical transformation 
$$ \alpha^{(\pm)}_{k\sigma}=u^{\pm}_k c^A_{k\sigma}+ 
u^{\mp}_k c^B_{k\sigma}, 
\eqno{(2)}
$$
with
$$ 
u^\pm_k=\sqrt{ \frac{1}{2} \left( 1\pm\frac{Jm_s}{2E_k} \right) },
\eqno{(3)}
$$ 
where $ c_{k\sigma}^{A(B)} $ is the Bloch transform 
of $ c_{i\sigma}^{A(B)} $ 
and $E_k=\sqrt{ {\epsilon_k}^2+{\left(Jm_s/2\right)}^2 }$ 
stands for the quasi-particle energy, and the wave vector
$k$ runs over reduced Brillouin zone. 
Making use of such a transformation, 
the Hamiltonian (1) 
is rewritten in the form : 
$$
H=H_{0}+H_{1}, 
$$
where
$$
H_0=\sum_{k\sigma}(E_k \alpha^{\left( + \right) \dagger}_{k\sigma}
\alpha^{\left( + \right) }_{k\sigma} 
 -E_k \alpha^{\left( - \right )\dagger}_{k\sigma}
 \alpha^{\left( - \right) }_{k\sigma}) 
                -Jm_c\sum_i ( S_i^{Az}-S_i^{Bz} ) 
\eqno{(4)}
$$
and 
$$
H_1= J\sum_i\{ ( m_s-S_i^{Az} )( m_c-\sigma_i^{Az})
              +( m_s-S_i^{Bz} )( m_c-\sigma_i^{Bz} ) \}
$$ 
$$
   +J\sum_{q}\{
    \sigma_q^{A+}S_{-q}^{A-}+\sigma_q^{A-}S_{-q}^{A+}
   +\sigma_q^{B+}S_{-q}^{B-}+\sigma_q^{B-}S_{-q}^{B+}\}, 
\eqno{(5)}
$$
where $\sigma_q$ and $S_q$ are the Bloch representations 
of spin operators of conduction and localized spins, respectively. 
The simple mean-field treatment, which neglects $H_1$, 
produces the correct ground state of the large-$S$ 
limit, providing the appropriate basis 
for further analysis. 
In order to study the effects brought about by the fluctuation $H_1$, 
we introduce the Green's function of localized spins, 
$\langle S^{A(B)+}_{q};S^{A-}_{-q} \rangle$, 
which satisfies the following equations of motion : 
$$
 z\langle S_q^{A+};S_{-q}^{A-} \rangle
 = 2\langle S_0^{Az} \rangle
 -J \sum_{q'}[ \langle \sigma_{q'}^{Az} S_{q-q'}^{A+};
 S_{-q}^{A-} \rangle
              +\langle S_{q-q'}^{Az} \sigma_{q'}^{A+};
 S_{-q}^{A-} \rangle ], 
\eqno{(6-a)}
$$
$$
 z\langle S_q^{B+};S_{-q}^{A-} \rangle
 =  -J \sum_{q'}[ \langle \sigma_{q'}^{Bz} S_{q-q'}^{B+};
 S_{-q}^{A-} \rangle
                  +\langle S_{q-q'}^{Bz} \sigma_{q'}^{B+};
 S_{-q}^{A-} \rangle ]. 
\eqno{(6-b)}
$$
Then the decoupling approximations such as 
$$
 \langle \sigma^{z} S^{+}; S^{-} \rangle
\longrightarrow
 \langle \sigma^{z} \rangle \langle S^{+}; S^{-} \rangle,
$$
$$
 \langle S^{z} \sigma^{+}; S^{-} \rangle
\longrightarrow
 \langle S^{z} \rangle \langle \sigma^{+}; S^{-} \rangle,
$$
are employed for the higher-order Green's functions, 
assuming the SDW state for carriers and N\'{e}el state 
for localized spins, respectively. 
This procedure is similar to the conventional Tyablikov method for 
the Heisenberg model. 
Thus we obtain 
$$
(z-Jm_c) \langle S_q^{A+}; S_{-q}^{A-} \rangle
=2m_s+Jm_s\sum_k
[ u_k^+u_{k+q}^+\langle \alpha_{k\uparrow}^{(+)\dagger}
                        \alpha_{k+q\downarrow}^{(-)}
               ;S_{-q}^{A-} \rangle
$$
$$
+ u_k^-u_{k+q}^-\langle \alpha_{k\uparrow}^{(+)\dagger}
                        \alpha_{k+q\downarrow}^{(-)}
               ;S_{-q}^{A-} \rangle ], 
\eqno{(7-a)}
$$
$$
(z+Jm_c) \langle S_q^{B+}; S_{-q}^{A-} \rangle
=Jm_s\sum_k
[ u_k^-u_{k+q}^-\langle \alpha_{k\uparrow}^{(+)\dagger}
                        \alpha_{k+q\downarrow}^{(-)}
               ;S_{-q}^{A-} \rangle
$$
$$
+ u_k^+u_{k+q}^+\langle \alpha_{k\uparrow}^{(+)\dagger}
                        \alpha_{k+q\downarrow}^{(-)}
               ;S_{-q}^{A-} \rangle ]. 
\eqno{(7-b)}
$$
In the above expression, the contributions from Green's functions 
of the type 
$\langle \alpha_{k\uparrow}^{(\pm)\dagger}a\alpha_{k+q\downarrow}^{(\pm)} ;S^{A-} \rangle$ 
are neglected, because the effects of 
the particle-particle and hole-hole scattering are sufficiently small 
at the weak coupling and low temperatures. 
Employing similar approximations, the following equations 
are derived for the Green's functions appearing on the r.h.s. 
of eqs.(7-a,b) \\
$
( z \pm E_k \pm E_{k+q} ) \langle 
\alpha_{k\uparrow}^{(\pm)\dagger}
\alpha_{k+q\downarrow}^{(\mp)}; S_{-q}^{A-} \rangle
=
$
$$ 
\frac{J}{2N}\{1-f(E_k)-f(E_{k+q})\}
[ u_k^\pm u_{k+q}^\pm \langle S_q^{A+};S_{-q}^{A-} \rangle
+ u_k^\mp u_{k+q}^\mp \langle S_q^{B+};S_{-q}^{A-} \rangle ], 
\eqno{(8)}
$$
where $f(E_k)$ is the Fermi distribution function.
From this closed set of equations, 
we obtain the explicit expression for the spin Green's function \\
$
{\langle}S^{A+}_{q};S^{A-}_{-q}{\rangle}=
$
$$
\frac{2m_s\left[
   \{ 1-\frac{{J}^2 m_s}{2} \phi_q(z) \} z
-\frac{{J}^2 m_s}{2}
\{ \chi_0^{AA}(0)+\chi_0^{AB}(0)-\chi^{AA}_q(z) \} 
             \right]}
        {
   {\{ 1-\frac{{J}^2 m_s}{2} \phi_q(z) \}}^2 z^2
  -{( \frac{{J}^2 m_s}{2} )}^2 
  [
  {\{ \chi_0^{AA}(0)+\chi_0^{AB}(0)-\chi^{AA}_q(z) \}}^2
  - {\{\chi^{AB}_{q}(z) \}}^2 
     ]
              }
\eqno{(9)}
$$
with
$$
   \phi_q(z)=\frac{2}{N}\sum_{k}
   \frac{u^{-2}_{k}u^{-2}_{k+q}-u^{+2}_{k}u^{+2}_{k+q}}
{z^{2}-(E_{k+q}+E_{k})^{2}}
\eqno{(10-a)}
$$
$$
   \chi^{AA}_q(z)=\frac{2}{N}\sum_{k}
   \frac{u^{-2}_{k}u^{-2}_{k+q}+u^{+2}_{k}u^{+2}_{k+q}}
   {z^2-(E_{k+q}+E_{k})^2}(E_{k+q}+E_{k})
\eqno{(10-b)}
$$
$$
   \chi^{AB}_q(z)=\frac{2}{N}\sum_{k}
   \frac{2u^-_k u^-_{k+q} u^+_{k} u^+_{k+q}}
   {z^2-(E_{k+q}+E_{k})^2}
   (E_{k+q}+E_{k}),
\eqno{(10-c)}
$$
where use has been made of the relation 
$$
m_c=-\frac{Jm_{s}}{2}\{\chi_0^{AA}(0)+\chi_0^{AB}(0)\}.
\eqno{(11)}
$$
Replacing $z$ in eq.(9) by $\omega +i\delta$, the selfconsistent
equation for the staggered magnetization $m_{s}$ is given 
by 
$$
 \frac{1}{2}-{m_{s}}= -\frac{1}{\pi} \frac{2}{N}\sum_{q}
{\int}^{\infty}_{-\infty}d{\omega}n(\omega) 
\mbox{Im}{\langle}S^{A+}_{q};S^{A-}_{-q}{\rangle}, 
\eqno{(12)}
$$
where $n(\omega)$ is the Bose distribution function 
$(e^{\beta\omega}-1)^{-1}$ and ${\beta}=T^{-1}$. 

This Green's function (9) has two types of poles corresponding to 
the Stoner and spin-wave excitations, respectively. 
Since the energy band of spin-wave excitations characterized by 
the energy scale $2Jm_c$ 
always lies lower than 
the continuum of the Stoner excitations 
 $ E_{k+q}+E_k $ which have a gap $2Jm_s$ 
in the whole momentum space, 
in the weak coupling regime the spin-wave dispersion is well
approximated in the form 
$$
\Delta(q)=\frac{{J}^2 m_s}{2} 
\frac{
\sqrt{ {\{ \chi^{AA}_0+\chi^{AB}_0-\chi^{AA}_q \}}^2-
{\{{\chi}^{AB}_{q}\}}
^{2}}}
{1-{J}^{2}m_{s}{\phi}_{q} /2 }, 
\eqno{(13)}
$$
by replacing 
$\chi \left( z \right) $ and $\phi \left( z \right) $ 
by their values at $z=0$. 
Our result $\left( 13 \right)$ clearly shows 
the presence of the infinitesimal spin excitation 
at q=0 
which is expected from the Goldstone theorem. 
We note that in the single-site approximation such as 
Gutzwiller and slave-boson methods, 
there remains a finite spin excitation gap even in AFO phase. 
Since the spectral intensities of Stoner excitations 
are sufficiently small 
compared to those of spin-wave excitations 
in the weak couling regime, 
in the following analysis we 
neglect the contributions from the Stoner excitations. 

 Let us first consider 
the asymptotic form of this dispersion at zero-temperature 
in the weak-coupling limit 
$J\rightarrow0$. 
In this limit, we can show that $\chi^{AA}_0$ and $\chi^{AB}_0$ 
diverge as $\ln J$, 
while $\chi^{AA}_q$ and $\chi^{AB}_q$($q\neq 0$) 
tend to a constant value. 
Therefore the equation (13) becomes independent of q in the limit 
$J\rightarrow0$ with a jump at $q=0$. 
This means that the results of the present theory 
at zero-temperature reduce to 
those of the mean-field theory in the weak-coupling limit, even
although the effects of quantum fluctuation 
are included in the present decoupling scheme. 
This asymptotic behavior is independent of the dimensionality 
of the system 
and consistent with our previous work of 1D PAM \cite{sh}. 

For general couplings, 
equation (13) has been solved numerically, and the results 
are shown in Fig.1: Here, considering the weak coupling case, 
we use $m_{s}=1/2$ and neglect deviation from it. 
The spin-wave dispersion strongly depends on the strength of 
Kondo coupling. 
It is found that the rapid increase in small-q region is gradually 
relaxed as $J$ becomes large. \\
\begin{center}
\psbox[scale=0.35]{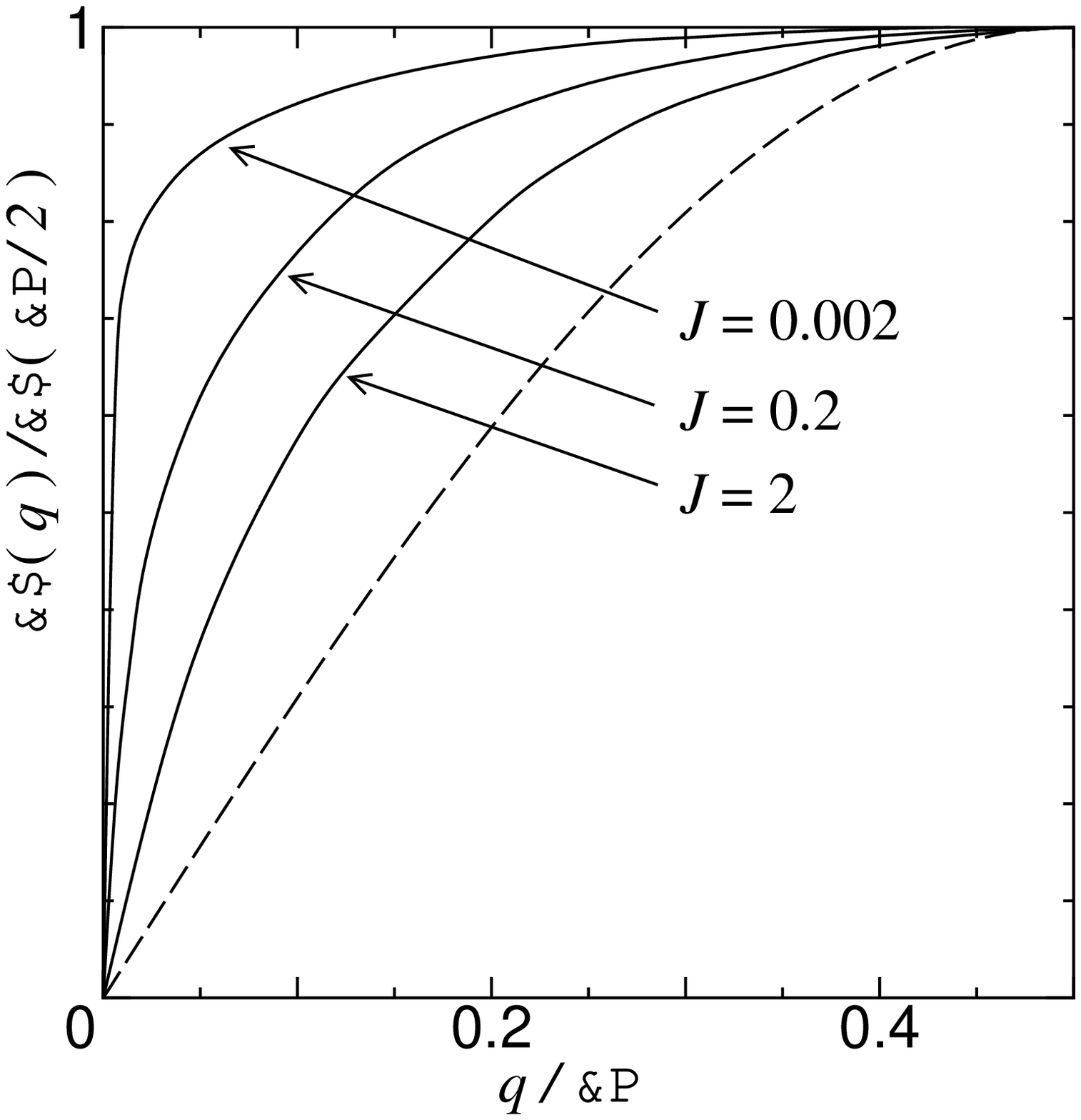}
\begin{quote}
\begin{footnotesize}
Fig.1. \hspace{6pt} 
Spin-wave dispersion, $\Delta \left( \vec{q} \right)$
 / $\Delta \left( \vec{\pi}/2 \right)$,
 as a function of $\vec{q}=\left( q,q,q \right) $
 for various Kondo coupling.
 The broken line represents the result in Heisenberg antiferromagnet 
; $\sqrt{1-\left( \cos\left( q_{x} \right) +
\cos\left( q_{y} \right) + \cos\left( q_{z} \right) \right)^{2}/9}$. \\
\end{footnotesize}
\end{quote}
\end{center}

 The linear expansion in 
$q$ for eq.(13) yields the spin-wave velocity 
$$
  v_{s}=0.088J\sqrt{ \ln{ \left( \frac{1}{J} \right) }},
\eqno{(14)}
$$
for 3D KLM in the weak-coupling region.
Similar results are also obtained for the lower dimensional
 systems : 
$v_{s}/J\sqrt{ \ln { \left( 1/J \right) }}=$
0.135 for 2D and 0.138 for 1D. 
Once the spin-wave velocity is determined, we can analytically
calculate the specific heat and parallel susceptibility 
in the low-temperature limit \cite{mat}\cite{app} as follows : 
$$
C_{v}=\frac{4\sqrt{3}\pi}{5} \left( \frac{T}{v_s} \right) ^3, 
\eqno{(15)}
$$
$$
 \chi_{//}=\frac{(\mu_B g)^2}{3} \frac{T^2}{v_s^3}.
\eqno{(16)}
$$


We can obtain the N\'{e}el temperature $T_{\rm N}$ 
by linearizing eq.(12) with respect to $ m_s $. 
Thus the equation for $T_{\rm N}$ is given by 
$$
 \frac{T_{\rm N}}{{J}^{2}}=\frac{\chi_{\rm N(0)}}{4}
\left[ \frac{2}{N}\sum_{q}\frac{2\chi_{\rm N}(0)+\chi_{\rm N}(q)}
{\chi_{\rm N}(0)-\chi_{\rm N}(q)} \right]^{-1}
\eqno{(17)}
$$
where 
$$
\chi_{\rm N}(q)=\frac{2}{N}\sum_{k}\frac{{1}-f_{\rm N}(\epsilon_{k})
-f_{\rm N}(\epsilon_{k+q})}{\epsilon_{k}+\epsilon_{k+q}}
\eqno{(18)}
$$
and $f_{\rm N}(\epsilon_{k})$ is the Fermi distribution function 
$(e^{\beta\epsilon_k}+1)^{-1}$ at $T=T_{\rm N}$. 
Equation (17) has been solved numerically with respect to $T_{\rm N}$ 
for various Kondo couplings $J$, and the results 
are shown in Fig.2. 
It is found that the N\'{e}el temperature shows a monotonic increase 
with increasing $J$. \\
\begin{center}
\psbox[scale=0.35]{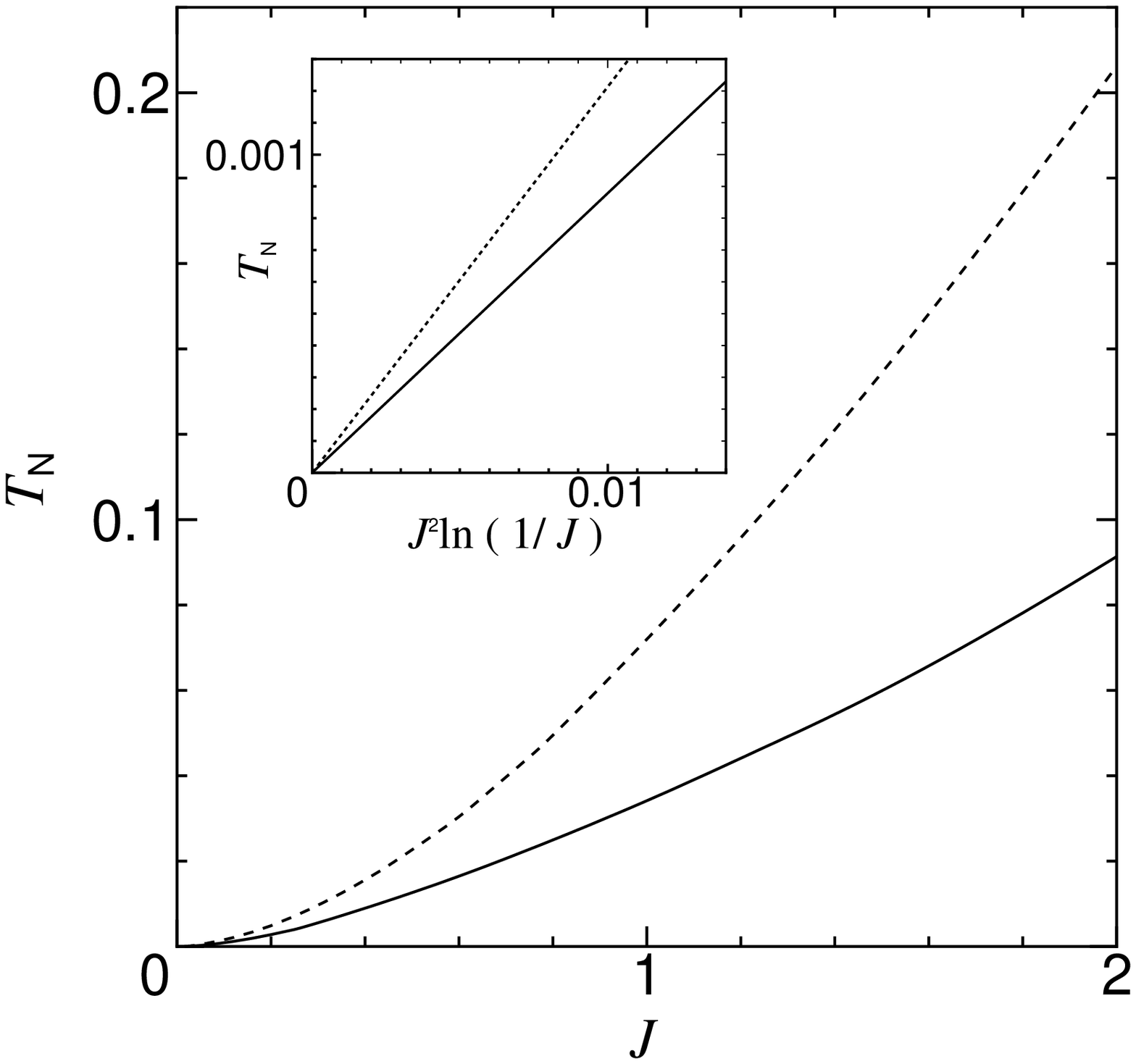}
\begin{quote}
\begin{footnotesize}
Fig.2. \hspace{6pt} N\'{e}el temperature $T_{\rm N}$ as a function
 of Kondo coupling. The solid and broken lines represent 
the present and the mean-field results, respectively.
 The inset shows that the N\'{e}el temperature
 is linear to ${J}^{2}\ln{(1/J)}$ in the weak-coupling region. 
\end{footnotesize}
\end{quote}
\end{center}

From the asymptotic behavior 
shown in the inset of Fig.2, we obtain the expression of $T_{\rm N}$ 
in the weak-coupling limit as follows : 
$$
T_{\rm N}= 0.087{J}^2\ln{\left(\frac{1}{J}\right)}. 
\eqno{(19)}
$$
The mean-field theory also predicts the 
$J$-dependence similar to (19) in the limit $ J \rightarrow 0$. 
However $T_{\rm N}$ in our theory is smaller than 
that in the mean-field theory by a factor 0.71. 
The deviation from 1 is due to the quantum fluctuations 
between localized spins. 
This is to be compared with the corresponding ratio 
0.66 in Heisenberg antiferromagnet.

It should be pointed out that the ratio of the spin-wave velocity 
and the N\'{e}el temperature $v_s/T_{\rm N}$ is enhanced as the 
Kondo coupling becomes small. 
This is because the long-range nature 
of effective interactions between localized spins 
suppresses collective excitations at low temperatures. 

In summary, the spin excitations of the AFO phase in 
half-filled KLM were studied by means of 
the decoupling approximation for spin Green's function. 
The spin-wave spectrum and the N\'{e}el temperature were
 calculated as a function of Kondo 
coupling, and their asymptotic behaviors in the weak-coupling limit 
were discussed in detail. 
It was shown that the collective spin excitations 
in weak-coupling KLM are considerably 
suppressed in comparison with those in the 
Heisenberg antiferromagnet, reflecting the long-range nature 
of effective interactions between localized spins. 

\begin{center}
 {\bf Acknowledgements}
\end{center}
 The authors are grateful to Dr.\ F. Suzuki for 
helpful comments and a careful reading of the manuscript.

\vspace*{5mm}

\begin{thebibliography}{99}
\bibitem{app}
It seems that these formulas could be applied 
if $T<v_{s}$. However this criterion overestimates 
the region in which these low-temperature approximations 
valid. The correct criterion is derived by 
comparing the contribution 
from the collective excitations 
given by eqs. (15) and (16) with that 
from the local excitations given by the mean-field theory. 
Thus the crossover temperature 
(below which the collective excitations 
dominantly govern the thermodynamic properties of the system) 
can be estimated as $T^{*}=\left[\ln (v_{s}/T_{\rm N})\right]T_{\rm N}$.
\end{thebibliography}
\end{document}